\documentclass[prl,twocolumn,showpacs]{revtex4}
\usepackage{graphicx}
\newcommand{\be}{\begin{equation}}
\newcommand{\ee}{\end{equation}}
\newcommand{\bea}{\begin{eqnarray}}
\newcommand{\eea}{\end{eqnarray}}

\newcommand{\tr}{{\rm Tr}}

\begin{document}
\title{Minimally entangled typical quantum states at finite temperature}
\author{Steven R. White}
\affiliation{Department of Physics and Astronomy, University of California,
 Irvine CA 92697, USA}
\date{\today}

\begin{abstract}
We introduce a class of states, called minimally entangled typical thermal
states (METTS), designed to resemble a typical state of a quantum system at
finite temperature with a bias towards classical (minimally entangled)
properties.  These states reveal in an intuitive way properties such as
short-range order which may be hidden in correlation functions.  An algorithm
is presented which, when used with the density matrix renormalization group
(DMRG), is faster by a factor of $10^3 - 10^{10}$ than previous heat-bath
approaches for thermally averaged quantities.  

\end{abstract}
\pacs{75.10.Jm, 75.40.Mg}
\maketitle

What is a typical wavefunction of a quantum system at a finite temperature?
The fundamental proposition of statistical mechanics is that the density matrix
of a system at inverse temperature $\beta$ with Hamiltonian $H$ is
$\rho=\exp(-\beta H)$. One can regard $\rho=\exp(-\beta H)$ as arising from 
several different
physical situations: from an ensemble average of pure states, from the long
time average of one system, from quantum mechanical entanglement with a heat
bath which produces mixed states, or from some combination of these effects.
The resulting predictions of statistical mechanics depend only on $\rho$.
On the other hand, statistical mechanics
is an idealization; a real physical system has a specific history and
environment which may favor thinking about it in one way over another.  Here we
will focus on the ensemble-of-pure-states point of view.  We have in mind
equilibrating the system with weak coupling to a heat bath, and then moving the
heat bath far from the system, removing any couplings.  From this viewpoint,
our question is a natural one.    In this paper we propose a set of idealized
states which we argue are useful to think of as ``typical'', and whose ensemble
generates $\rho=\exp(-\beta H)$.  In addition, the algorithm we introduce to
generate them provides a substantially more
efficient route to determining finite temperature properties of lattice models
when using diagonalization, density matrix renormalization group
(DMRG)\cite{dmrg}, and tensor product wavefunction approaches\cite{twodim,Mera}.

What do we mean by typical? We mean that there is a set of states
$\{|\phi(i) \rangle\}$ with unnormalized probabilities $P(i)$, from which we can
select states.  To reproduce statistical mechanics, we require
\begin{equation}
\sum_i P(i) |\phi(i) \rangle \langle \phi(i)| =  e^{-\beta H} .
\label{Pphirho}
\end{equation}
Then the expectation value of any Hermitian operator $A$ can be determined by
an unweighted average of $\langle \phi(i) | A | \phi(i)\rangle$, with each
$|\phi(i) \rangle$ chosen at random according to $P(i)$.
We also impose looser criteria based on physics:  that one can imagine some
physical thermalization process which might generate the $\{|\phi(i) \rangle\}$,
and that the $\{|\phi(i) \rangle\}$ do not exhibit special ``atypical'' physical characteristics.
We {\it do not} require that every
state in the Hilbert space be included in the $\{|\phi(i) \rangle\}$.

For classical systems on a lattice, the only reasonable 
typical states are the
{\it classical product states}  (CPS), 
$|i\rangle = \prod_{{\rm sites }\ \ell}|i_\ell \rangle$, where $i_\ell$ labels
the states of a site.
For example, for an Ising model a CPS is a spin configuration,
e.g. $|i\rangle = |\uparrow \downarrow \downarrow \uparrow \ldots \rangle$.
These states are often generated numerically and provide an intuitive understanding of a system's
properties which would be difficult
to obtain from the system's density matrix.  
For quantum spin systems, one can also generate CPS, but these are not typical
wavefunctions.  For example, at temperature $T=0$, the typical wavefunction
should be the ground state, which is generally not a CPS.

The energy eigenvalues $E_s$ and eigenstates $|s\rangle$ satisfy
$\rho = \sum_s e^{-\beta E_s} |s\rangle\langle s| $ and thus Eq. (\ref{Pphirho}).
However, they should not be thought of as typical states.
Schr\"odinger called this idea ``altogether wrong" and 
``irreconcilable with the very foundations of quantum mechanics"\cite{schrodinger}.
For a large system, excluding very low temperature, equilibration
processes  do not drive the system to any single eigenstate. 
Any such process would take an exponentially long time (in the number of particles $N$)
because of the exponentially small energy level spacing.
The eigenstates are also
exponentially sensitive to uncertainties in the Hamiltonian. 
Nevertheless, more recent introductions to statistical mechanics than 
Schr\"odinger's often
give the impression (sometimes without explicitly saying so) that
the typical thermal wavefunction is an eigenstate of the Hamiltonian\cite{feynman}.

It is easy to construct other states satisfying Eq. (\ref{Pphirho}).
Let $\{|i\rangle\}$ be any complete orthonormal basis of the system.
Define the normalized (but not orthogonal) set of ``typical'' states
\begin{equation}
|\phi(i)\rangle =  P(i)^{-1/2} \exp(-\beta H/2) |i\rangle,
\end{equation}
where 
\begin{equation}
P(i) \equiv \langle i | \exp(-\beta H) |i\rangle = \tr\{\rho |i\rangle\langle i| \}.
\label{Peq}
\end{equation}
Note that the partition function is given by $Z = \tr \rho = \sum_i P(i)$.
We see Eq. (\ref{Pphirho}) immediately follows, and
\begin{equation}
\langle A \rangle \equiv \frac{1}{Z} \tr \{ \rho A \} = \sum_i \frac{P(i)}{Z} 
\langle \phi(i) | A | \phi(i)\rangle .
\label{Aeq}
\end{equation}
Note that similar results are obtained if the states $\{|i\rangle\}$ are not orthonormal,
provided there exists a positive set of weights $p(i)$ such that
$\sum_i p(i) |i \rangle \langle i| =  1$,
and similarly for a continuous distribution of states.

The energy eigenstates can serve as the set $|i\rangle$, in which case 
$|\phi(i) \rangle = |i\rangle$.
Another choice is to select the $\{|i\rangle\}$ as  random
normalized vectors in the Hilbert space,  selected using the Haar measure;
this might be considered a mathematically 
natural definition of typical states.  Both
of these approaches are intractable computationally except on the smallest systems.  
Exact energy eigenstates would be unsuitable
even for a  Lanczos approach because of the 
small  level spacings--a full diagonalization of $H$
would  be required.  These choices are also poor from a physical point
of view. In a broken symmetry phase, these states would tend to be
highly non-classical superpositions of many states with different values of the
order parameter.  If the system consisted of two widely separated
noninteracting subsystems, the random vector approach would give highly
entangled states of the two subsystems.  These choices ignore decoherence
effects, which tend to eliminate highly-non-classical entanglement.

We take as our favored ``typical'' states $\{|\phi(i) \rangle\}$ the ones with
the least entanglement, generated by taking $\{|i\rangle\}$ to be a CPS.
By entanglement we mean we
consider dividing the system into two parts, say by taking a dividing plane,
and calculating the von Neumann entanglement entropy $S$ between the two parts.
A CPS has $S=0$ for any division.
At nonzero $\beta$ we expect the resulting $|\phi(i) \rangle$ to
have minimal entropy within this general class of states, and so we call them
minimally entangled typical thermal states (METTS). 
METTS have a number of nice properties.  A METTS for a system with noninteracting
subsystems is a product of METTS for the subsystems.  For systems with long-range order,
METTS break symmetries,
choosing an order parameter at random.  Even for systems without broken symmetries,
METTS reveal underlying short-range order.

We give two approaches to the calculation of METTS.  The 
{\it ancilla method}\cite{verstraetethermal,feiguinthermal}  uses a set of
auxilliary sites acting as a heat bath, and
 has been used to simulate finite temperatures using time dependent
DMRG methods\cite{tdmrg}.   It's adaptation to produce METTS
resembles a highly idealized physical thermalization process.
The {\it pure state method} does not use a heat bath and  is more efficient.

To describe
the ancilla method\cite{verstraetethermal},   let us take the system $A$ to 
be composed of $N$ spins with $S=1/2$.
The heat bath $B$ is also composed of $N$ $S=1/2$ spins (called ancilla), 
and we pair up the spins in $A$ and $B$---for a 1D system, think of a ladder.
Label the sites by $\ell$ and let $|i_\ell\rangle_A$ label the local states of the system site at $\ell$,
and similarly for $B$. 
The initial unnormalized pure state of $A+B$ which describes infinite temperature is
\begin{equation}
|\psi(\beta=0)\rangle = \sum_{i_1}\ldots \sum_{i_N} |i_1\rangle_A |i_1\rangle_B 
\ldots |i_N\rangle_A |i_N\rangle_B
\end{equation}
This state is a product   of site-ancilla pair states, with each pair  maximally
entangled. If one traces out the ancilla from $|\psi\rangle\langle \psi|$, one obtains the
infinite temperature density matrix $1$. The Hamiltonian of $A+B$ is that
 of $A$ alone:  there are no $A-B$ or $B-B$ terms. Let
\begin{equation}
|\psi(\beta)\rangle = \exp(-\beta H/2) |\psi(0)\rangle.
\end{equation}
We find $Tr_B{|\psi(\beta)\rangle \langle \psi(\beta)|} = \exp(-\beta H)$. Alternatively,
one can measure any operator $O$ of $A$ as 
$\langle O \rangle= \langle \psi(\beta)| O |\psi(\beta)\rangle$.  
The calculation of $|\psi(\beta)\rangle$
is easily performed using imaginary time-dependent DMRG\cite{tdmrg}, 
with initial state $|\psi(\beta=0)\rangle$.

 To obtain a pure state for $A$ from the entangled state of $A+B$, we perform a 
{\it physical measurement}  of all the spins of $B$.    A physical measurement projects the
wavefunction into one eigenstate of the measured operator, with the appropriate probability.
Specifically, to measure one
particular spin in the $z$ direction, compute 
$P(\uparrow) = \langle S_z \rangle + \frac{1}{2}$,  
let $P(\downarrow)= 1 -P(\uparrow)$, and set
\begin{equation}
| \psi \rangle \to \left\{ \begin{array}{ll}
P(\uparrow)^{-\frac{1}{2}}|\uparrow\rangle \langle \uparrow |\  \psi
\rangle & {\rm prob }\  P(\uparrow)\\
P(\downarrow)^{-\frac{1}{2}}|\downarrow\rangle \langle \downarrow |\
\psi \rangle & {\rm prob } \  P(\downarrow) \end{array} \right.
\end{equation}
Note that we get the same probability distribution whether
we measure the sites sequentially or jointly all at once, but sequentially is much more convenient numerically,
taking one half-sweep in DMRG\cite{ueda2005}.
We are free to measure each  spin with respect to any axis,  all the same or different, randomly or
predetermined.
The probability of the final CPS $|i\rangle_B$  is
 given by Eq. (\ref{Peq}).
The measurement puts the combined system  into the product state
\begin{equation}
P(i)^{-1/2} |i\rangle_B \langle i |_B \psi(\beta)\rangle = |i\rangle_B |\phi(i)\rangle_A
\end{equation}
At this point one can ignore $B$.
Note that the initial perfect entanglement takes the place of
the coupling one would have in a real thermalization process.

In the pure-state method, we start with any CPS $|i\rangle$, and  
apply $\exp(-\beta H/2)$.  We then physically measure a new CPS $|i'\rangle$ from this state, 
and apply $\exp(-\beta H/2)$ to it, etc.
We call one iteration a ``thermal step''. This process
resembles Monte Carlo, but with the quantum measurement process taking the
place of the usual spin flips.  
The set of  METTS are a fixed point of this process:
consider an infinite ensemble of such systems, initially with $|i\rangle$ distributed with
probability $P(i)$. Then by Eq. (\ref{Aeq}), the ensemble of  $|\phi(i)\rangle$  correctly reproduces all
thermodynamic measurements, so a set of $|i'\rangle$ determined from it is correctly distributed with probability $P(i')$. 

To study METTS in more detail, we consider the one dimensional $S=\frac{1}{2}$ Heisenberg model, with Hamiltonian
\begin{equation}
H = \sum_\ell \vec S_\ell \cdot \vec S_{\ell+1} ,
\end{equation}
and with open boundary conditions. We implement the algorithms using
time-dependent DMRG with a second order breakup and with a time step of 0.05.
The physical measurements generating the $|i\rangle$ were done at
different random orientations for each spin and thermal time step. 

\begin{figure}[!tb]
\begin{center}
\includegraphics*[width=0.65\hsize,scale=1.0]{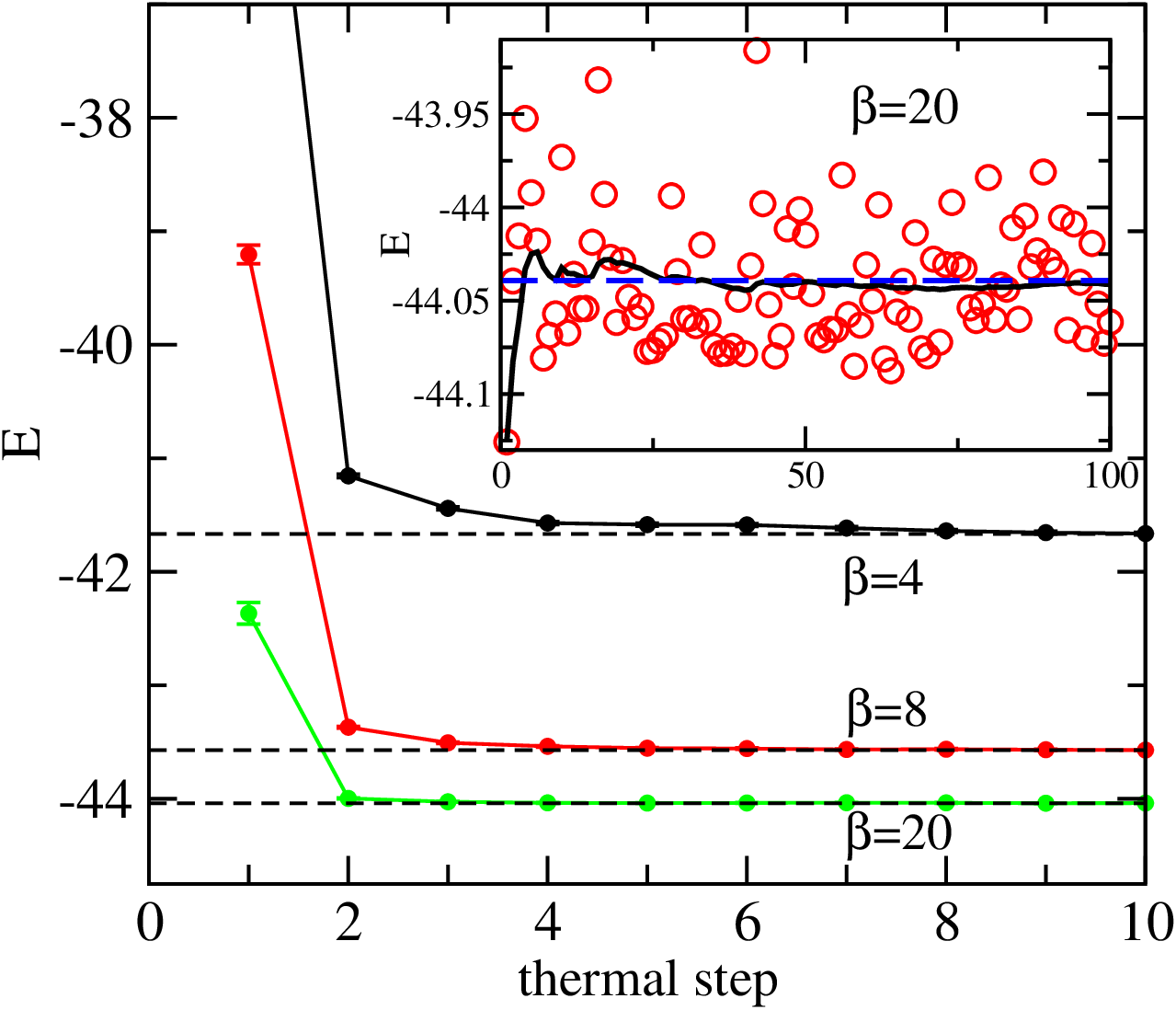}
\caption{Energy of a 100 site Heisenberg chain at various temperatures.  In the main figure, 
the dashed lines label the results of the ancilla approach, considered exact.
The symbols are derived from repeated use of the pure-state method, 
with each use
starting from a new completely random state $|i\rangle$ 
and proceeding 10 thermal steps.  For
a fixed thermal step, we averaged over the ensemble.
The inset shows one long pure-state calculation.  The open
circles show energies of individual METTS, while the solid line shows the
moving average.  The dashed line shows the ancilla result.
}
\label{fig:Emc}
\end{center}
\end{figure}

In Fig. 1 we show  that the two algorithms give the same (numerically exact) 
results for the energy on a 100 site Heisenberg chain.  The main figure shows
that for the pure-state method, one reaches the equilibrium distribution 
for the energy very precisely after 5-10 thermal steps, starting from a random
configuration.  This suggests that
the thermal-step autocorrelation time
(similar to a Monte Carlo autocorrelation time)
is very short, the key to efficient sampling.  In practical calculations
(inset) one does one long run with many thermal steps, discarding the first
results as being a warmup, say about 10 steps.  Averaging over only 100 METTS
obtained in $N_\tau = 100$ thermal steps ($+10$ for the warmup), we obtain the
total energy to a relative accuracy of about $10^{-5}$.  The fluctuations in
the total energy are quite small; one can obtain reasonable results with only
one METTS.

With DMRG, particularly for low temperatures and modest accuracies, the pure-state METTS method is much faster
than the ancilla method for obtaining thermal averages.  (For averages, there is no point in generating
METTS with the ancilla method, since averages can be measured directly.)  Suppose for a specified accuracy
 a system requires $m_0$ states
per block for a $T=0$ standard DMRG calculation.  In pure-state METTS we solve the imaginary-time-dependent 
Schr\"odinger equation from 0 to $\beta/2$. We find that for pure-state METTS, the $m$ required starts
at 1 for small imaginary time and saturates to $m_0$ for very large imaginary times, which are only
needed for large $\beta$.  The calculation time scales as $N m_0^3 \beta N_\tau$, where $N$ is the number of sites.
In the ancilla method, in the limit of low temperatures, the heat bath and the system both independently encode
the ground state, 
as a product state but with their sites intermingled.  This means that DMRG requires $m_0^2$ states, and
the calculation time scales as $N m_0^6 \beta$, bigger by a factor of $m_0^3/N_\tau$ compared to the pure-state
METTS approach. Typical values of $m_0$ are $50-5000$ for systems ranging from simple 1D spin chains to 2D clusters with
width $8-10$.  Consequently, taking $N_\tau = 10-100$, the pure-state METTS method is faster by
a factor of $10^3 - 10^{10}$.

\begin{figure}[t]
\begin{center}
\includegraphics*[width=0.8\hsize,scale=1.0]{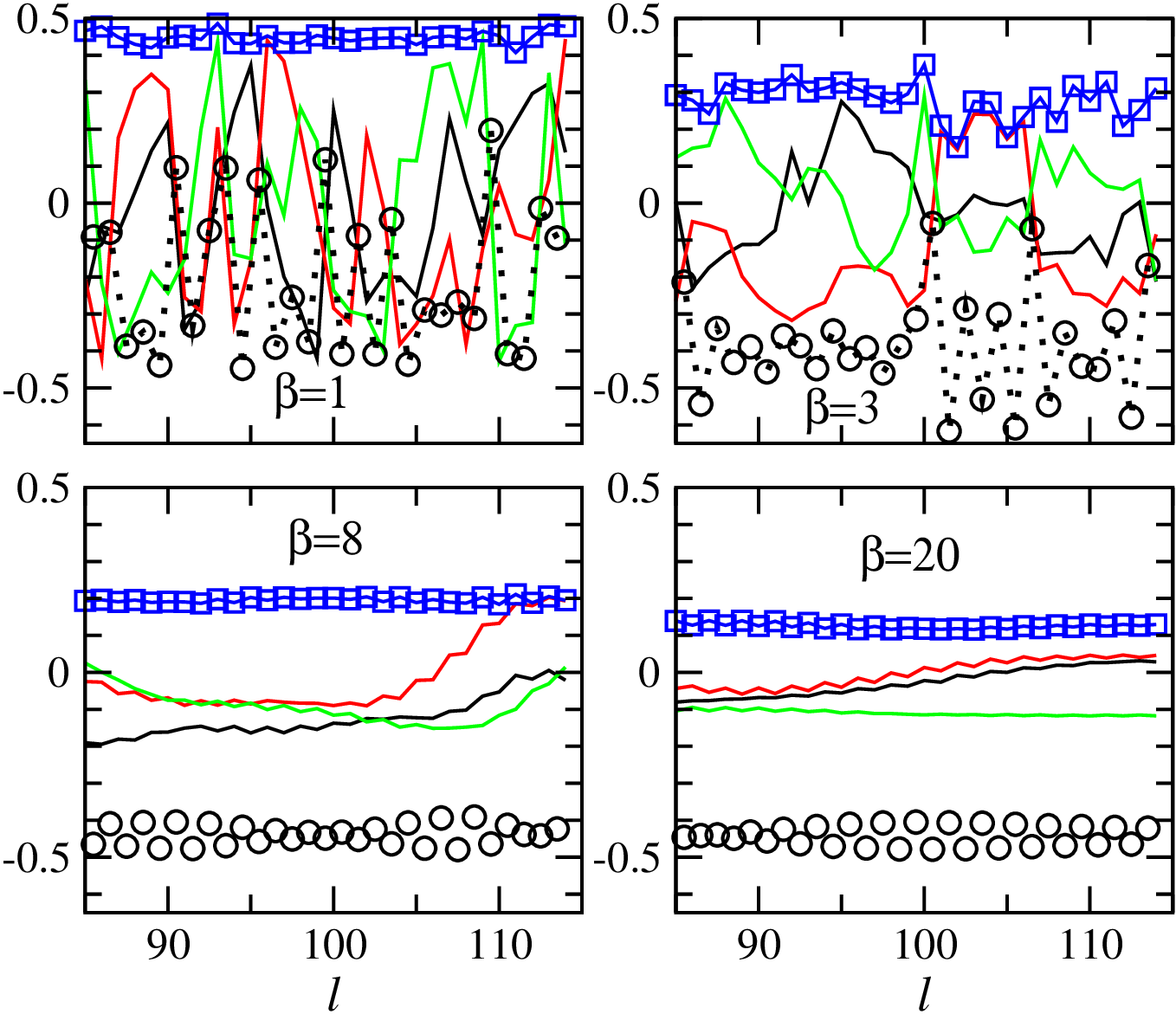}
\caption{Properties  of METTS for a 200 site Heisenberg chain, central 30 sites.  
Each panel shows properties of a single METTS generated 
for the indicated temperature (the METTS in the different panels are unrelated).  
The three solid
lines (red, black, green) without symbols show 
$(-1)^\ell \langle S^\alpha \rangle$, for $\alpha=x,y,z$ (which line is $x$, etc., 
is arbitrary).  The open squares at the top show $\cal C_\ell$.
The open circles show
$\langle \vec S \cdot \vec S \rangle$ on each bond.
}
\label{fig:typical}
\end{center}
\end{figure}

In Fig. 2 we show properties of some METTS for a Heisenberg chain.  All the
measurements show substantial randomness, which diminishes at lower
temperatures as the METTS approach the ground state. Since the model is
antiferromagnetic, we multiply the spin measurements by $(-1)^\ell$ to make
twisting of the antiferromagnetic order more apparent. For example, for $\beta=8$,
pronounced twisting is visible near $\ell=105-110$. The values of $\langle \vec
S \cdot \vec S \rangle$ show an increase in dimerization in the same region.
Similar twisting and dimerization is visible at $\beta=3$ near  $\ell=100-108$.
We know that at finite temperature, the system has a finite spin-spin
correlation length; this could come about (we imagine) via random twisting of
the spin order, by regions with strong dimerization, or some combination.  In
these METTS both effects occur, with  twisting being somewhat more pronounced.
The open squares measure ${\cal C}_\ell \equiv (\langle S^x_\ell
\rangle^2+\langle S^y_\ell \rangle^2+\langle S^z_\ell \rangle^2)^{1/2}$.  If
the measurements included ensemble averaging, $\cal C_\ell$ would always be
zero.  Instead, it measures how classical a spin is---how entangled it is
within the METTS.  For an isolated $S=1/2$ in any pure state, ${\cal
C}_\ell=1/2$.  Any total $S=0$ wavefunction would give ${\cal C}_\ell=0$ for
every $\ell$.  The METTS are biased to be as classical as possible, which makes
${\cal C}_\ell$ meaningful.  It is surprising how little variation there is in
${\cal C}_\ell$ from site to site.

\begin{figure}[!t]
\begin{center}
\includegraphics*[width=0.5\hsize,scale=1.0]{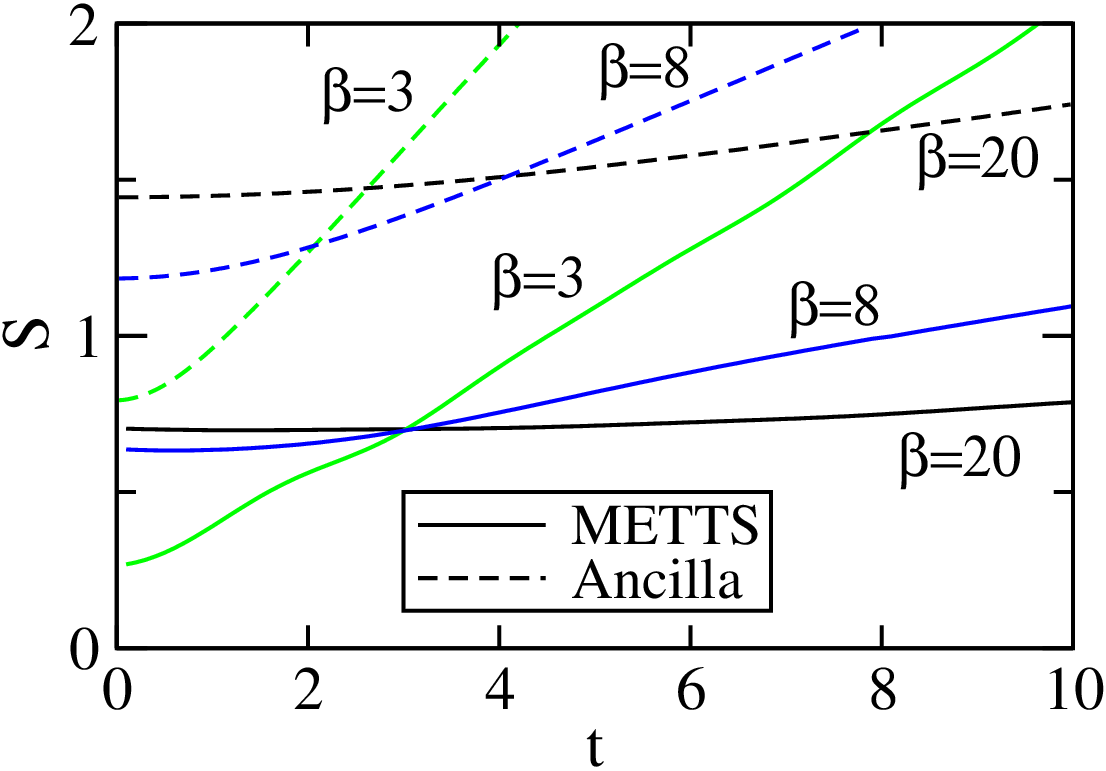}
\caption{Entanglement entropy at the center bond of a 40 site Heisenberg chain
as a function of real time.
Each solid line is for a single METTS,
while the dashed lines are for an ancilla system with 40 sites and 40 ancilla.
}
\label{fig:spec_wt}
\end{center}
\end{figure}

The METTS can be evolved in real time (say with real-time DMRG).
The ensemble averages of METTS states are time independent, but the METTS
themselves are not. Much as a single particle with a narrowly peaked
wavefunction would spread out in time, METTS evolve to states with much higher
entanglement entropy.  In Fig. 3 we show the growth of $S$ with time for
several different temperatures. In the higher temperature cases, the entropy
starts smaller but grows more rapidly.  The same effect is seen in the results
for an ancilla system, for which the typical entropy is roughly twice that of
the METTS, in agreement with the behavior of $m$, with $m \sim \exp(S)$.  The
behavior of the entropy as a function of time determines the effectiveness of
real-time DMRG to calculate finite temperature spectral
functions\cite{barthel}.  Our results show that METTS will be able to reach
longer times than the ancilla approach\cite{barthel}.

The rapid growth of entanglement with time for METTS raises the 
question of whether METTS (at $t=0$) 
really are ``typical'' wavefunctions of real systems. The answer is very likely no, 
typical wavefunctions have more entanglement than METTS, with eventual entanglement
growth limited by decoherence. One can evolve an
ensemble of METTS to some fixed time $t$; the resulting set of states
also satisfy Eq. (\ref{Pphirho}), would exhibit more entanglement, and thus could be considered as being
more realistic physically.  However, the METTS themselves are more
useful computationally.

We briefly note several other approaches to finite temperatures.
A quite different (but powerful) finite temperature DMRG approach for infinite,
translationally invariant 1D systems is transfer matrix DMRG\cite{tmdmrg,Sirker}. 
More closely related to our work are two approaches adapted for Lanczos
calculations\cite{ftlm,mclm} and one recent DMRG approach\cite{sota}.  Both Lanczos
approaches start with a random vector.  One utilizes completeness properties of Lanzcos (Krylov)
expansions to produce an approximation to canonical ensemble results\cite{ftlm}.   The other approach
produces microcanonical results by minimizing $(H-\lambda)^2$, obtaining not a single eigenstate
but a superposition of many which is narrow in energy\cite{mclm}.
The DMRG approach\cite{sota} starts with a random vector chosen from the (incomplete) DMRG basis
and uses a regulated polynomial expansion to apply $\exp(-\beta H/2)$.
None of the approaches utilize physical measurement or a CPS starting state $|i\rangle$ chosen with probability $P(i)$.
We believe
METTS has a more rigorous foundation, e.g. not dependent on any completeness
properties of the Lanczos or DMRG basis, and applicable to any temperature.
METTS also provide very useful intuition about the nature of the system at
finite temperature.

We acknowledge very helpful discussions with F. Verstraete, A.L. Chernyshev,  G. Vidal, 
T. Barthel, and G. Chan.
We acknowledge  support from the NSF under grant DMR-0605444.

\end{document}